\documentclass[aps,prl,reprint]{revtex4-1}

\usepackage{graphicx}
\usepackage{subfigure}
\usepackage{color}
\usepackage{amsmath}
\usepackage[linktocpage,colorlinks=true,linkcolor=blue,citecolor=blue,breaklinks=true,urlcolor=blue]{hyperref}
\usepackage{verbatim}
\usepackage{breakcites}
\usepackage{wrapfig}
\usepackage{soul}

\begin{document}

\title{Intrinsic Free Energy in Active Nematics}

\author{Sumesh P. Thampi}
\affiliation{The Rudolf Peierls Centre for Theoretical Physics, 1 Keble Road, Oxford, OX1 3NP, UK}

\author{Amin Doostmohammadi}
\affiliation{The Rudolf Peierls Centre for Theoretical Physics, 1 Keble Road, Oxford, OX1 3NP, UK}

\author{Ramin Golestanian}
\affiliation{The Rudolf Peierls Centre for Theoretical Physics, 1 Keble Road, Oxford, OX1 3NP, UK}

\author{Julia M. Yeomans}
\affiliation{The Rudolf Peierls Centre for Theoretical Physics, 1 Keble Road, Oxford, OX1 3NP, UK}

\newcommand{\ty}[1]{\textcolor{red}{{#1}} }
\newcommand{\ar}[1]{\textcolor{blue}{{#1}}}
\newcommand{\ad}[1]{\textcolor{cyan}{{#1}} }
\newcommand{\ju}[1]{\textcolor{green}{{#1}} }

\date{\today}

\begin{abstract}
\noindent
Basing our arguments on the theory of active liquid crystals, we demonstrate both analytically and numerically, that the activity can induce an effective free energy which enhances ordering in extensile systems of active rods and in contractile suspensions of active discs. We argue that this occurs because any ordering fluctuation is enhanced by the flow field it produces. A phase-diagram in the temperature-activity plane compares ordering due to a thermodynamic free energy to that resulting from the activity. We also demonstrate that activity can drive variations in concentration, but for a different physical reason that relies on the separation of hydrodynamic and diffusive timescales. 
\end{abstract}


\maketitle

\section{Introduction}
Suspensions of active particles such as catalytic colloids, bacteria, cells and cytoskeletal filaments exhibit fascinating dynamical behaviours, from clustering and swarming \cite{Dombrowski2004,Wolgemuth2008, Saha2014} to phase separation \cite{Shelley2007,Igor2013} and turbulent like flows \cite{Julia2012,Jorn2013}. These are examples of non-equilibrium states of matter as the individual constituent units continuously input energy and generate motion. Recent studies have uncovered a wealth of novel phenomena in these systems, reviewed in \cite{Sriram2010, Ganesh2011, Marchetti2013, Sriram2003,Joanny2005,Mahadevan2011,Chate2012,Dogic2012,Giomi2013,SumeshPRL2013,Giomi2014turb}.

Continuum models have demonstrated a promising capability in predicting salient characteristics of active systems such as collective motion and mesoscale turbulence. In particular continuum models based on the equations of nematic liquid crystals, known as active nematics, have been shown to successfully reproduce a number of experimental observations \cite{Dogic2012,Giomi2014turb, Giomi2013, SumeshPRL2013, Julia2012, Jorn2013}. However, in accounting for various mechanisms, these phenomenological models contain a large number of parameters that often makes it difficult to pinpoint the consequences of activity or to move towards quantitative comparisons to experiment.

The active nematic equations include terms that correspond to relaxation towards a free energy minimum describing the equilibrium configuration of the system \cite{Sriram2010,Marchetti2013}. Often this is the Landau de Gennes free energy pertinent to nematic liquid crystals, but free energy contributions that drive phase ordering of concentration or density have also been considered \cite{Mahadevan2011,Giomi2013,Baskaran2014}. Including a free energy arises naturally if the continuum models describing the non equilibrium state of active systems are derived by following the Onsager procedure as this is based on the assumption of small perturbations to an equilibrium state \cite{Marchetti2013}. Moreover, when active entities are elongated it seems reasonable to assume that they will adopt a nematic configuration that can be described by a Landau-de Gennes free energy. However the role of a free energy in describing eg spherical bacteria is not immediately apparent: here the nematic nature of the continuum equations is solely a consequence of the symmetry of the dipolar flow field produced by active particles.

Therefore in this paper we concentrate on the role of free energy and demonstrate how activity induced hydrodynamics competes with free energy driven dynamics. First we introduce the theoretical description of active nematics. This allows us to demonstrate that activity itself can act as an effective free energy that changes the degree of nematic ordering. Activity modifies the dynamics, through renormalising all of the terms in an effective Laudau--de Gennes free energy, in a way that depends on the coupling between the nematic ordering and the self-generated active flow.
We present physical arguments interpreting our results. A phase diagram in the activity--temperature plane (shown in Fig. \ref{fig:phase}) illustrates the interplay between activity and free energy driven ordering. This analysis allows us to (i) link active liquid crystal theory with purely fluid dynamical approaches, (ii) analyse the concept of effective temperature in the context of active nematics, and (iii) propose a reduced description for the dynamics of extensile active nematics, and hence to list the dimensionless variables most relevant to controlling active nematohydrodynamics, without free energy contributions.

\section{Equations of motion}
The local orientational order of an active nematic material is described using a tensor order parameter $\mathbf{Q} = \frac{q}{2} ( 3\mathbf{nn} - \mathbf{I})$, where $ \mathbf{n}$ is the director and $q$ the magnitude of the order. The local concentration of active particles is represented by $\phi$ and total density and velocity field are denoted by $\rho$ and $\mathbf{u}$, respectively. In the absence of any underlying bulk free energy, the coupled evolution equations that describe the hydrodynamics of active nematics are:
\begin{align}
\partial_t \rho + \partial_i (\rho u_i)&=0,
\label{eqn:cont}\\
\rho (\partial_t + u_k \partial_k) u_i &= \partial_j \Pi_{ij},
\label{eqn:ns}\\
(\partial_t + u_k \partial_k) Q_{ij} - S_{ij} &= \Gamma_Q H_{ij},
\label{eqn:lc}\\
\partial_{t}\phi+\partial_{i}(u_{i}\phi)&=\Gamma_{\phi} K_{\phi}\nabla^{2}\phi.
\label{eqn:conc}
\end{align}

The generalised advection term in Eq.~\ref{eqn:lc},
\begin{align}
S_{ij} = &(\lambda E_{ik} + \Omega_{ik})(Q_{kj} + \frac{\delta_{kj}}{3}) + (Q_{ik} + \frac{\delta_{ik}}{3})\nonumber\\
 &(\lambda E_{kj} - \Omega_{kj}) - 2 \lambda (Q_{ij} + \frac{\delta_{ij}}{3})(Q_{kl}\partial_k u_l),
 \label{eqn:cor}
\end{align}
describes the response of $\mathbf{Q}$ to velocity gradients. Here, $E_{ij} = (\partial_i u_j + \partial_j u_i)/2$ is the strain rate tensor and $\Omega_{ij} = (\partial_j u_i - \partial_i u_j)/2$ is the vorticity tensor. The alignment parameter $\lambda$ determines the collective response of active particles to a velocity gradient. Comparison to the Leslie-Ericksen formulation of nematic liquid crystals indicates that particles align (tumble) if $\lambda_1>1$ ($\lambda_1 < 1$) where $\lambda_1 = (3q+4)\lambda/9q$. This parameter also takes into account the shape of the particles: $\lambda>0$ and $\lambda<0$ correspond to rod-like and plate-like particles respectively \cite{Berisbook, Davide2007, Scott2009}. We denote the rotational diffusion coefficient of the orientation field as $\Gamma_Q$ and the molecular field $H_{ij}=K_Q \nabla^2 Q_{ij}$ where $K_Q$ is an elastic constant. $\Gamma_{\phi}$ is the  diffusivity of the concentration field and $K_{\phi}$ is related to the energy attributed to gradients in $\phi$. Note in particular that the governing equations do not assume the presence of any underlying bulk free energy.

The standard, passive liquid crystal contributions to the stress are the viscous stress, $\Pi_{ij}^{viscous} = 2 \eta E_{ij}$, and the elastic stress,
\begin{align}
\Pi_{ij}^{passive}&=-P\delta_{ij} + 2 \lambda(Q_{ij} + \delta_{ij}/3) (Q_{kl} H_{lk})\nonumber\\
&-\lambda H_{ik} (Q_{kj} + \delta_{kj}/3)  - \lambda (Q_{ik} + \delta_{ik}/3) H_{kj}\nonumber\\
&-K\partial_i Q_{kl}\partial_j Q_{kl} 
+ Q_{ik}H_{kj} - H_{ik} Q_{kj},
\end{align}
where $\eta$ is the viscosity and $P-\frac{K_Q}{2}(\partial_k Q_{ij})^2$ is the modified pressure.
The active contribution to the stress is a consequence of the dipolar nature of the forces exerted by active particles, $\Pi_{ij}^{active} = -\zeta \phi Q_{ij}$ \cite{Sriram2002}. Thus, any gradient in $\mathbf{Q}$ generates a flow field of strength that depends on the concentration $\phi$ and the activity coefficient, $\zeta$ with $\zeta>0$ for extensile and $\zeta<0$ for contractile systems. 

\begin{figure}
\subfigure[]{\includegraphics[trim = 240 40 240 30,clip,width=0.49\linewidth]{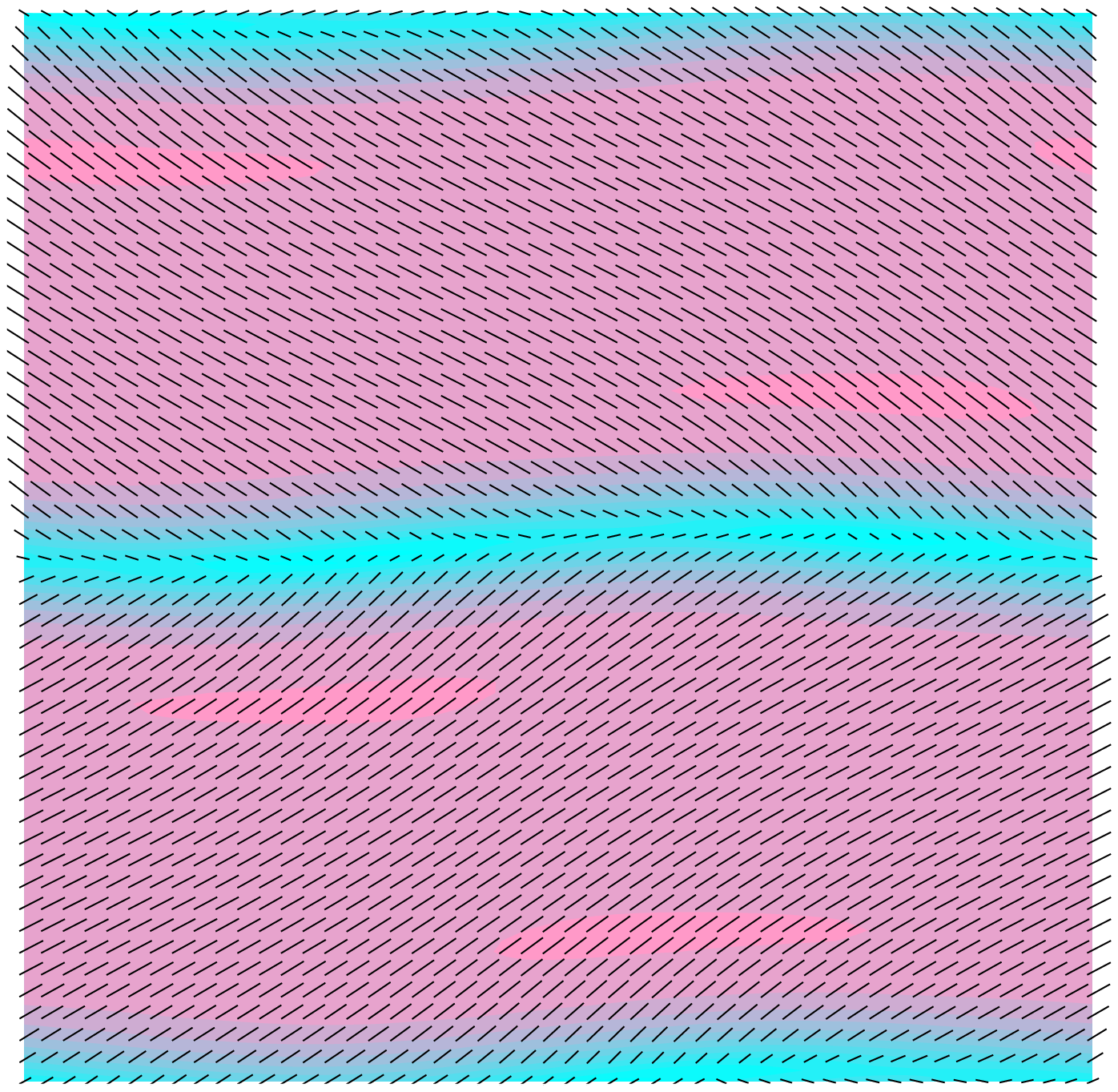}}
\subfigure[]{\includegraphics[trim = 240 40 240 30,clip,width=0.49\linewidth]{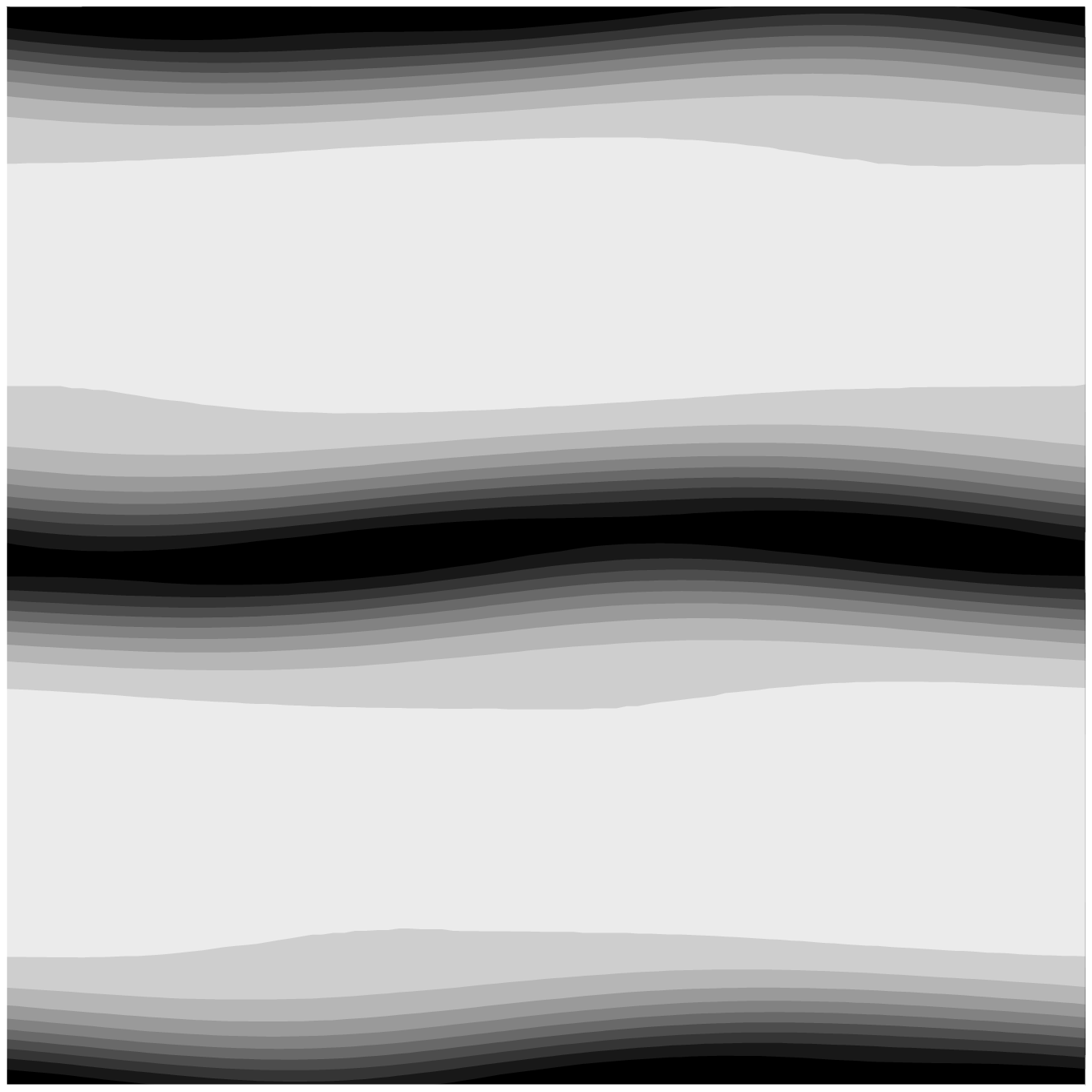}}
\subfigure[]{\includegraphics[trim = 270 50 263 30,clip,width=0.49\linewidth]{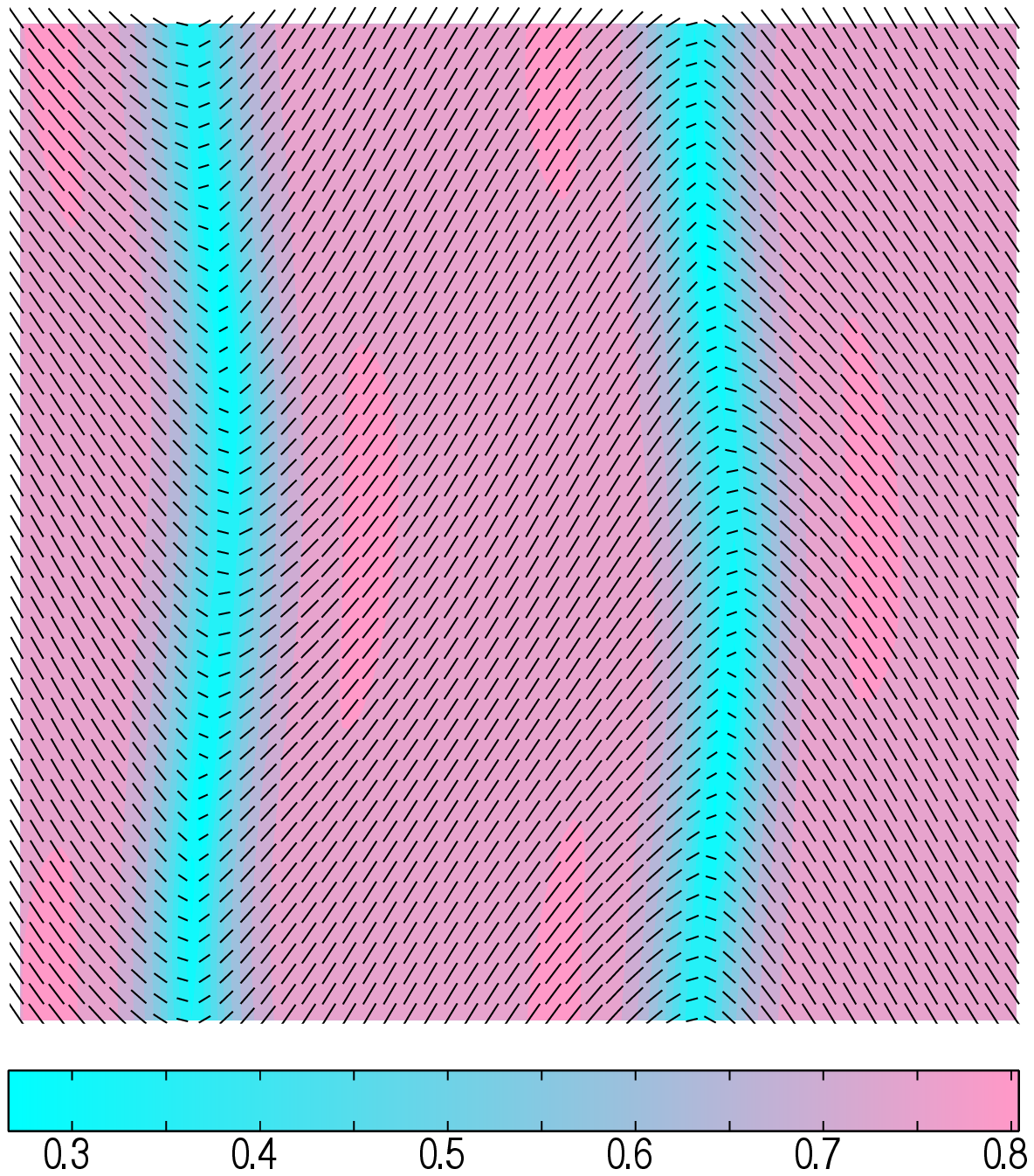}}
\subfigure[]{\includegraphics[trim = 270 50 263 30,clip,width=0.49\linewidth]{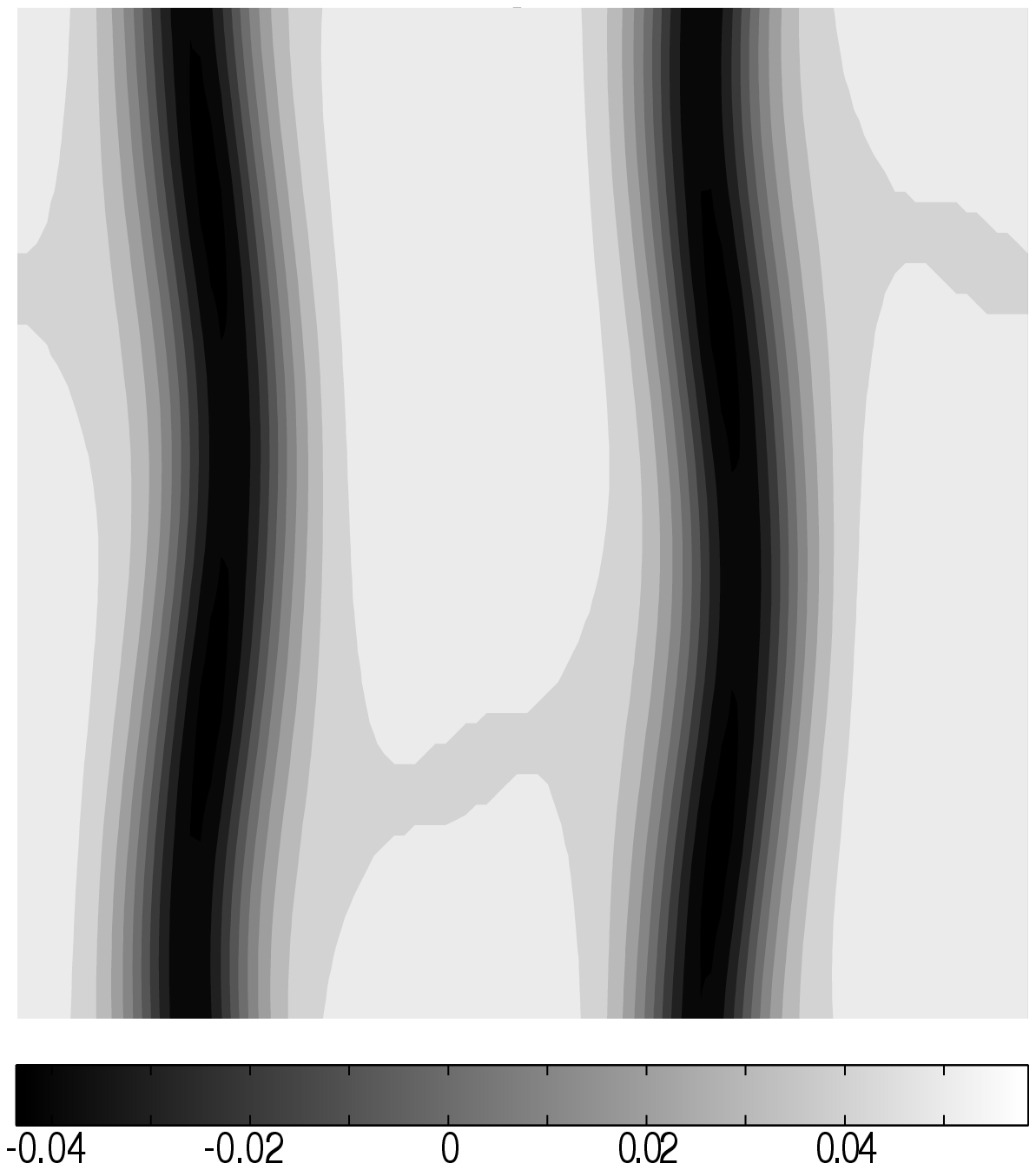}}
\caption{Walls in the director field obtained by solving the active nematohydrodynamic equations with zero bulk free energy. The system size is 100$\times$100. (a) Director field; the colour shading is the magnitude of the orientational order parameter. (b) Corresponding concentration field, $\delta\phi=\phi-\phi^{*}$ with $\phi^{*}=0$. (c), (d) Similar frames at a later time showing the switching behaviour of the walls. 
}
\label{fig:bands}
\end{figure}

\begin{figure*}
\begin{center}
\subfigure[]{\includegraphics[trim = 440 72 410 60,clip,width=0.31\linewidth]{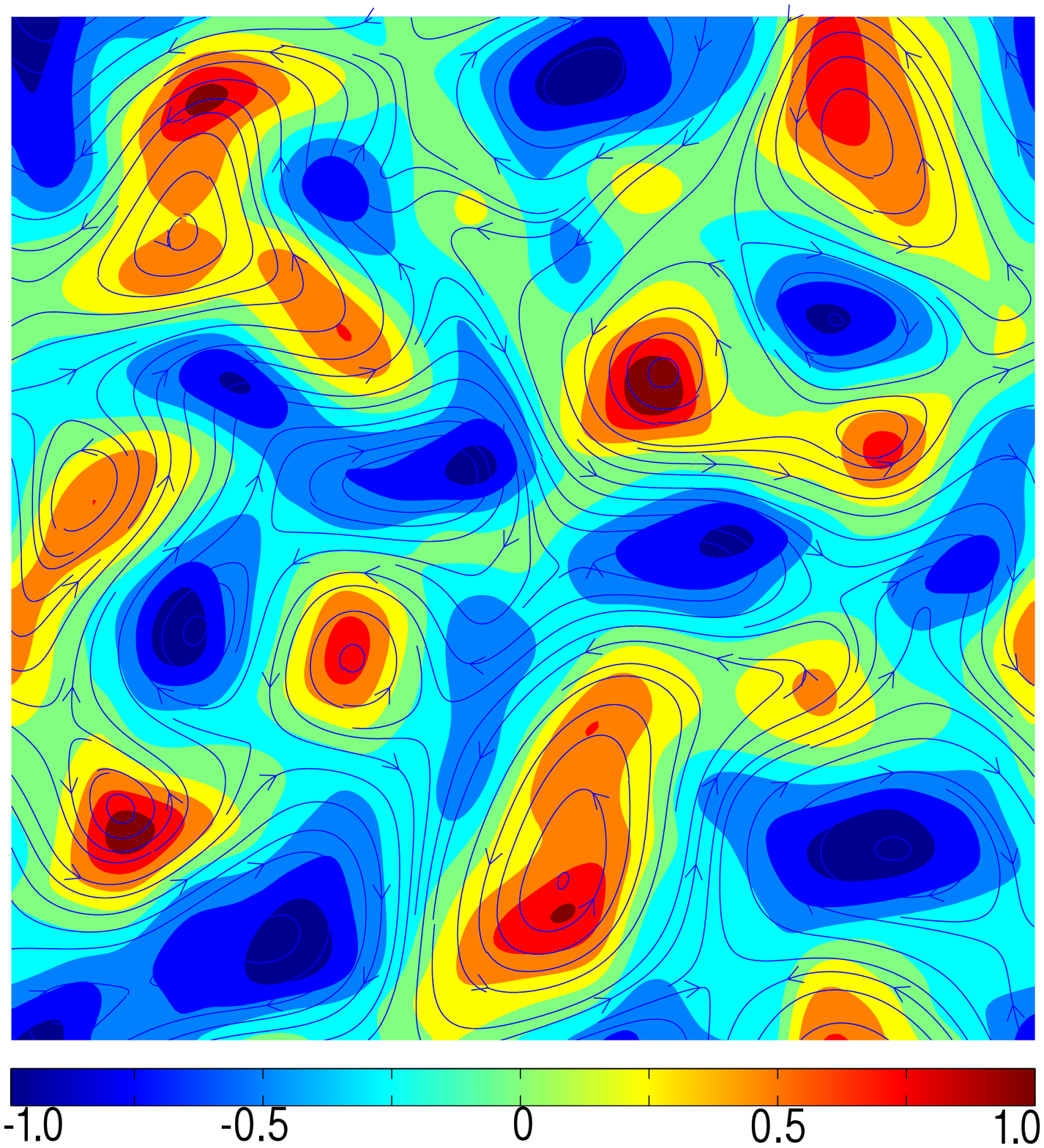}}
\subfigure[]{\includegraphics[trim = 440 72 410 60,clip,width=0.31\linewidth]{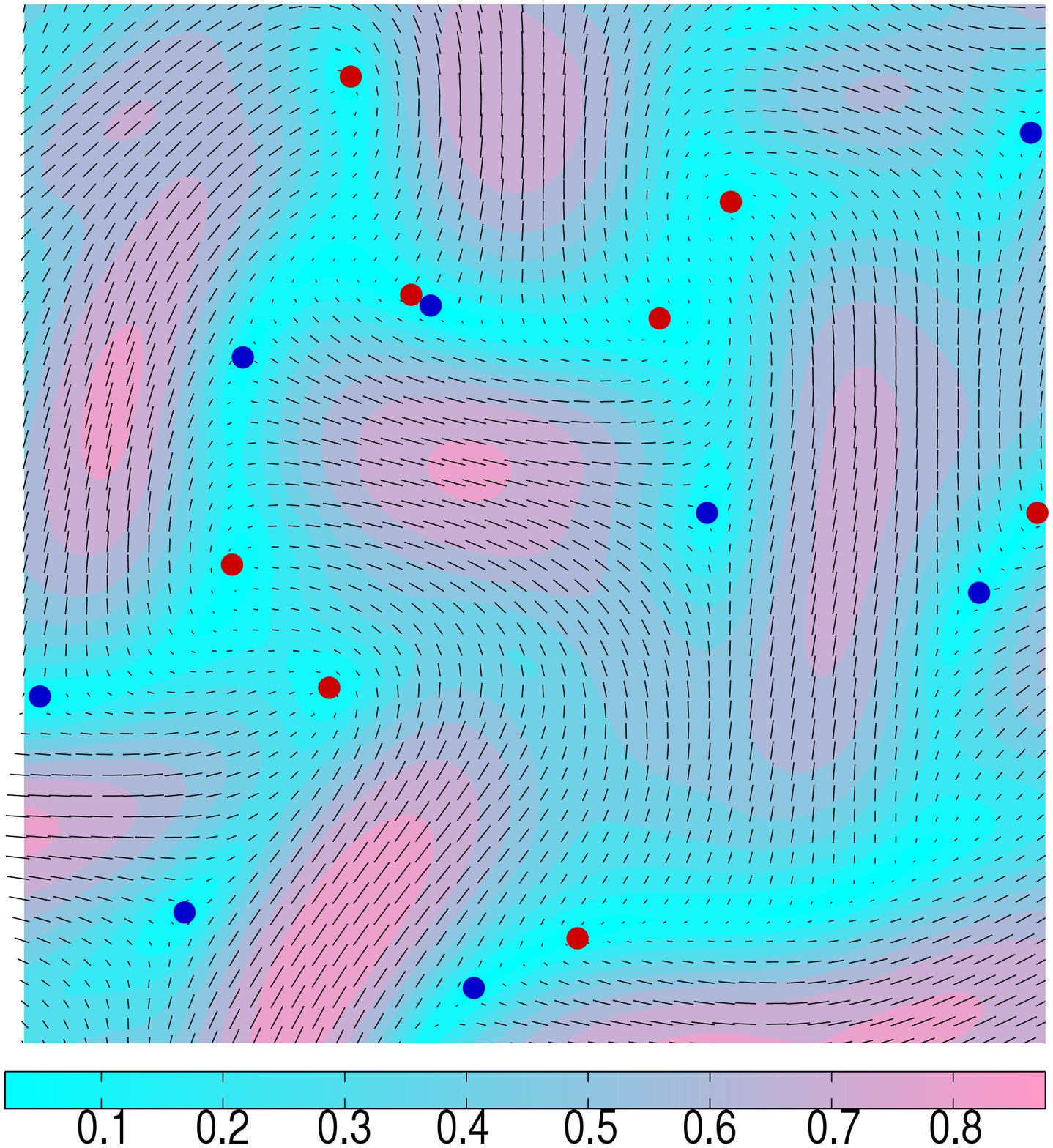}}
\subfigure[]{\includegraphics[trim = 440 72 410 60,clip,width=0.31\linewidth]{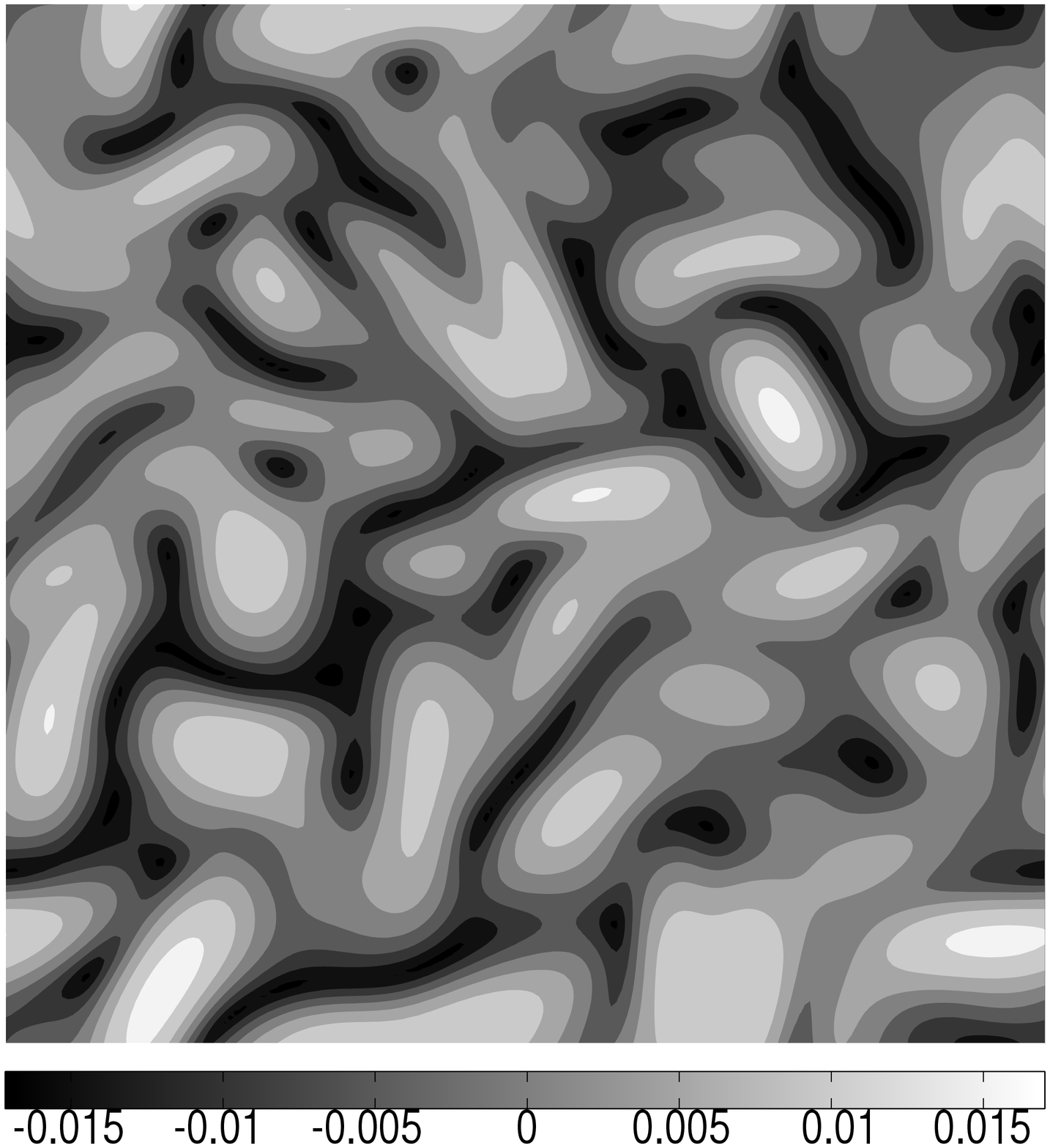}}
\end{center}
\caption{Active turbulence obtained by solving the active nematohydrodynamic equations with zero bulk free energy. The system size is 200$\times$200.
(a) Turbulent-like flow field. Continuous lines are stream lines and colour shading shows the vorticity  normalised by its maximum value. (Blue, clockwise; red, anticlockwise.) (b) Director field, colour shading shows the magnitude of the order parameter. Red and blue dots respectively represent the $\pm \frac{1}{2}$ topological defects. (c) Corresponding concentration field. Only the left bottom quarter of panels (a) and (c) is shown in panel (b).}
\label{fig:turb}
\end{figure*}

The equations of active nematohydrodynamics (Eqs.~\ref{eqn:cont}--\ref{eqn:conc}) are solved using a hybrid lattice Boltzmann method \cite{ourpta2014}. The parameters used in the simulations are $\Gamma_Q=0.34 $, $\Gamma_{\phi} = 1$, $K_Q=0.01$, $K_{\phi} = 1$, $\zeta=0.01$ and $\mu=2/3$, in lattice units, unless mentioned otherwise. Simulations were performed in a two-dimensional domain of size $100 \times 100$ for Fig.~\ref{fig:bands}, $200 \times 200$ for Fig.~\ref{fig:turb} and 
 $400 \times 400$ elsewhere.

\section{Activity-induced ordering and phase separation}
We first present results for the dynamic behaviour of an extensile active nematic. The simulations were started with zero order and uniform concentration in the entire system.
Fig.~\ref{fig:bands}(a) shows that, despite the lack of an ordering free energy, the system forms local nematic domains. These are separated by  \textit{walls}, lines of strong distortions in the director field, which are created because the uniform active nematic state is unstable to the flow fields induced by any fluctuation in the nematic order.  Panel (b) of Fig.~\ref{fig:bands} indicates that the concentration is slightly lower at the walls where the magnitude of the nematic order parameter is lower. The walls are unstable, they periodically deform, break up and change their orientation to re-form in a new configuration that is perpendicular to the previous configuration as shown in panels (c) and (d). 
This dynamics has been reported earlier as one characteristic feature of active nematics \cite{Mahadevan2011, Giomi2014}. The difference here is that we show that the behaviour persists even in the absence of any ordering free energy.

The band dynamics depicted in Fig.~\ref{fig:bands} is, however, a finite size effect. We find a transition to a turbulent state at higher activities and larger domain sizes (Fig. \ref{fig:turb}(a)), as reported earlier \cite{Giomi2014}.
 As evident from Fig. \ref{fig:turb}, in a domain twice as large as that in Fig.~\ref{fig:bands},  pairs of oppositely charged topological defects form at the walls. These move apart, driven by flow and elastic forces. When a pair of oppositely charged defects meet, they annihilate restoring nematic regions which can then undergo further instabilities. The resulting chaotic dynamics, now observed in several experiments and simulations, has been termed active turbulence  \cite{Dogic2012, Julia2012, Jorn2013, Giomi2013, SumeshPRL2013, Shi2013, MatthewPRL, Shelly2014,Giomi2014turb,Decamp2015}.  Fig.~\ref{fig:turb}(c) shows that the concentration field correlates with the nematic order in the active turbulent state, with lower concentration values corresponding to a lower value of the nematic order parameter.

Thus we have demonstrated numerically that the active nematohydrodynamic equations (1)--(4) produce characteristic features of active dynamics even in the absence of an underlying bulk free energy.
Activity has played two distinct roles here: (i) generating activity-induced nematic ordering and (ii) generating concentration gradients. We consider each in turn.

\section{Nematic ordering}
We now give analytical arguments to motivate the activity-induced nematic ordering. To do this, we neglect (i) the fluid inertia, which is negligible on the length scales of bacterial suspensions and cytoskeletal filaments, (ii) the elastic stress which generates back flow (in our simulations, the flow due to elastic stress is one to two orders of magnitude smaller than that due to activity), and (iii) any pressure gradients. Thus we may balance the viscous stress with the active stress and write the approximate local equation
\begin{align}
2 \eta E_{ij}  \approx \zeta \phi Q_{ij}.
 \label{eqn:EQ}
\end{align}
Considering this as an equality and simplifying the generalised co-rotational term the Eq.~(\ref{eqn:lc}) becomes
\begin{equation}
(\partial_t + u_k \partial_k) Q_{ij} - \Omega_{ik} Q_{kj} + Q_{ik} \Omega_{kj}= \Gamma_Q H_{ij},
\end{equation}
where
\begin{align}
&H_{ij}=K_Q \nabla^2 Q_{ij}+\frac{\lambda \zeta}{3 \eta} \phi Q_{ij} \nonumber \\
&+ \frac{\lambda \zeta}{ \eta} \phi \left[Q_{ik}Q_{kj} - Q_{pq}Q_{qp}\frac{\delta_{ij}}{3} \right]-  \frac{\lambda \zeta}{ \eta} \phi Q_{ij}Q_{pq}Q_{qp} .
\label{eqn:corsim}
\end{align}
The new terms that appear in the molecular potential correspond exactly to an effective Landau-deGennes free energy \cite{DeGennesBook}
\begin{align}
&\mathcal{F}_{A}=\frac{\lambda \zeta}{6 \eta \Gamma_Q}\phi Q_{ij} Q_{ji}  \nonumber \\
&- \frac{\lambda \zeta}{ 3\eta \Gamma_Q}\phi Q_{ij} Q_{jk} Q_{ki}+
\frac{\lambda \zeta}{4 \eta \Gamma_Q}\phi(Q_{ij} Q_{ji})^2. 
\label{eqn:effE}
\end{align}
Thus the active stress results in effective free energy terms, proportional to the alignment parameter $\lambda$ and the strength of the activity $\zeta$. These will change the magnitude of local nematic ordering.
\begin{figure}
\subfigure[]{\includegraphics[trim = 0 0 0 0, clip, width=0.30\linewidth]{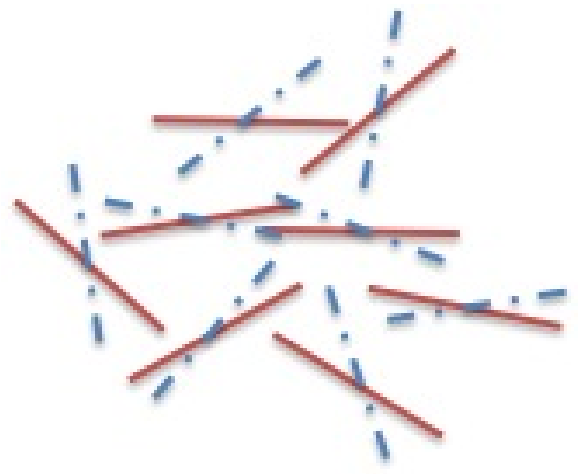}}
\subfigure[]{\includegraphics[trim = 0 0 0 0, clip, width=0.34\linewidth]{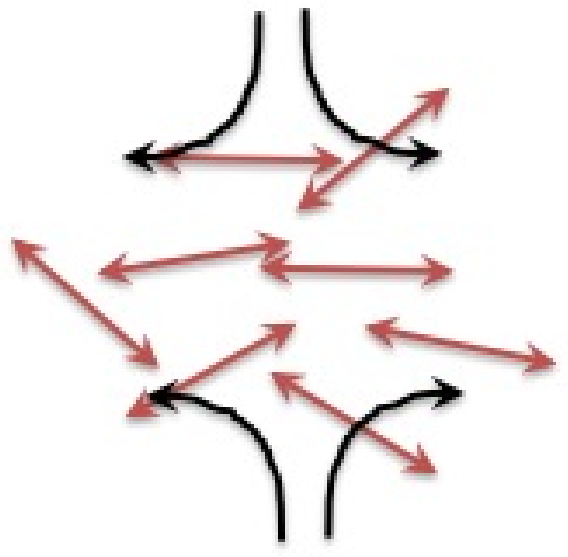}}
\subfigure[]{\includegraphics[trim = 0 0 0 0, clip, width=0.31\linewidth]{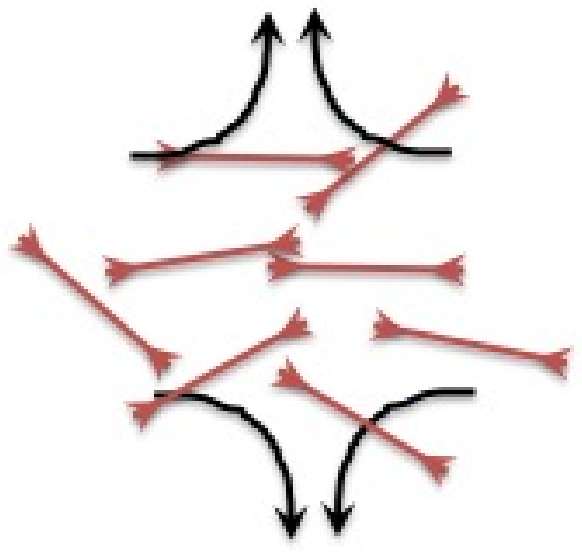}}
\caption{(a) In an isotropic configuration of rod like particles (dashed lines), fluctuations may induce local, temporary ordering (continuous lines). (b) Extensile particles generate flows with extensional axis along the instantaneous orientation which stabilises this ordering. (c) Contractile particles generate flows with compressional axis along the instantaneous orientation, which destabilises the ordering.}
\label{fig:mecha}
\end{figure}

We now provide the physical interpretation for this renormalisation of the free energy (Fig.~\ref{fig:mecha}). A rod ($\lambda>0$) attains a stable position in an extensional flow when it is aligned along the extensional axis. Starting from an isotropic configuration of elongated particles (Fig.~\ref{fig:mecha}(a)) consider a fluctuation that induces a small ordering, along a given direction. If the particles are active they generate dipolar flow fields, which enhance the instantaneous order for an extensile system or destabilise it in the contractile case. Thus the response of the $Q$ tensor to the extensional part of the flow generates an effective molecular potential and, in extensile systems, the hydrodynamics acts as an effective force favouring the alignment of the active rods.
For plate-like particles, $\lambda<0$, however, the stable position is along the compressional axis with plate orientation defined normal to the plate. Therefore now it is the contractile stress that favours ordering.
Since these mechanisms depend only on the extensional part of the flow field, they hold for both aligning and tumbling particles.

These arguments agree with the dependence of the effective free energy terms in Eq.~(\ref{eqn:effE}) on the product $\lambda \zeta$. Simulations showed, as expected, that for $\lambda \zeta<0$, a system initialised in the isotropic state with zero bulk free energy remained in this state, whereas for a system with $\lambda \zeta>0$ we observed active turbulence (Fig.~\ref{fig:turb}).

The nematically ordered regions in active systems are known to be hydrodynamically unstable \cite{Sriram2002, Marchetti2013}. This is again a consequence of the response of the active particles to self-generated flow fields. While we have argued that the extensional part of the active flow (that is proportional to $\lambda$ in $S_{ij}$) can enhance orientational ordering, the vortical part of the active flow plays the major role in destabilising the nematic state \cite{Madan2007, Scott2009}. Understanding active turbulence as a competition between these two mechanisms warrants further investigation.

\begin{figure}
\begin{center}
\includegraphics[trim = 0 0 20 0, clip, width=0.8\linewidth]{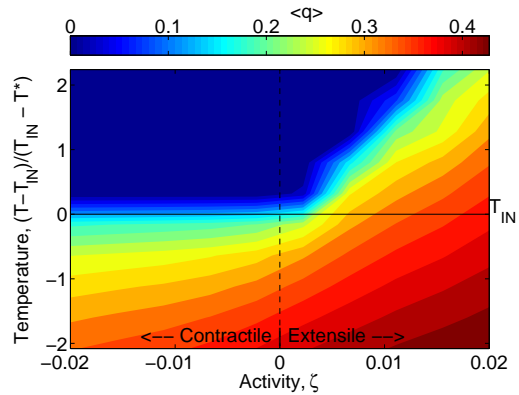}
\end{center}
\caption{Average value of the magnitude of the order parameter, $\langle q \rangle$, in a plane spanned by normalised temperature and activity for rod like particles ($\lambda=0.7$).}
\label{fig:phase}
\end{figure}

We now add the usual  thermodynamic, Landau de Gennes free energy terms 
\begin{equation}
\mathcal{F_{TD}}=\frac{A_0(T-T^*)}{2} Q_{ij} Q_{ji}  \nonumber \\
+\frac{B}{3} Q_{ij} Q_{jk} Q_{ki}+
\frac{C}{4}(Q_{ij} Q_{ji})^2
\end{equation}
into the equations of motion where they appear as an additional term  in the molecular field $H_{ij}$. $T_{IN}=T^{\ast}+\frac{B^2}{27A_0C}$ is the isotropic-nematic transition temperature in the passive liquid crystal. Restricting ourselves to systems which have a uniform concentration ($\phi = $constant) we capture the competition between the free energy-induced and activity-induced ordering by measuring the average magnitude of order parameter $\langle q \rangle$, independently changing $T$ and $\zeta$ in the simulations (Fig.~\ref{fig:phase}.)

First consider extensile systems. Below $T_{\textnormal{IN}}$, $\langle q \rangle$ increases with $\zeta$ due to activity induced ordering. The effective isotropic-nematic transition temperature $T_{\textnormal{IN}} (\zeta)$ increases approximately linearly with $\zeta$. In contractile systems, however, activity destroys nematic ordering. Thus $\langle q \rangle = 0$ even below $T=T_{\textnormal{IN}}$. This disordering is very small and cannot be seen clearly in Fig.~\ref{fig:phase} but the reduction in $\langle q \rangle$ with increased contractile activity is clear. 

\begin{figure}
\subfigure[]{\includegraphics[trim = 0 0 0 0, clip, width=0.48\linewidth]{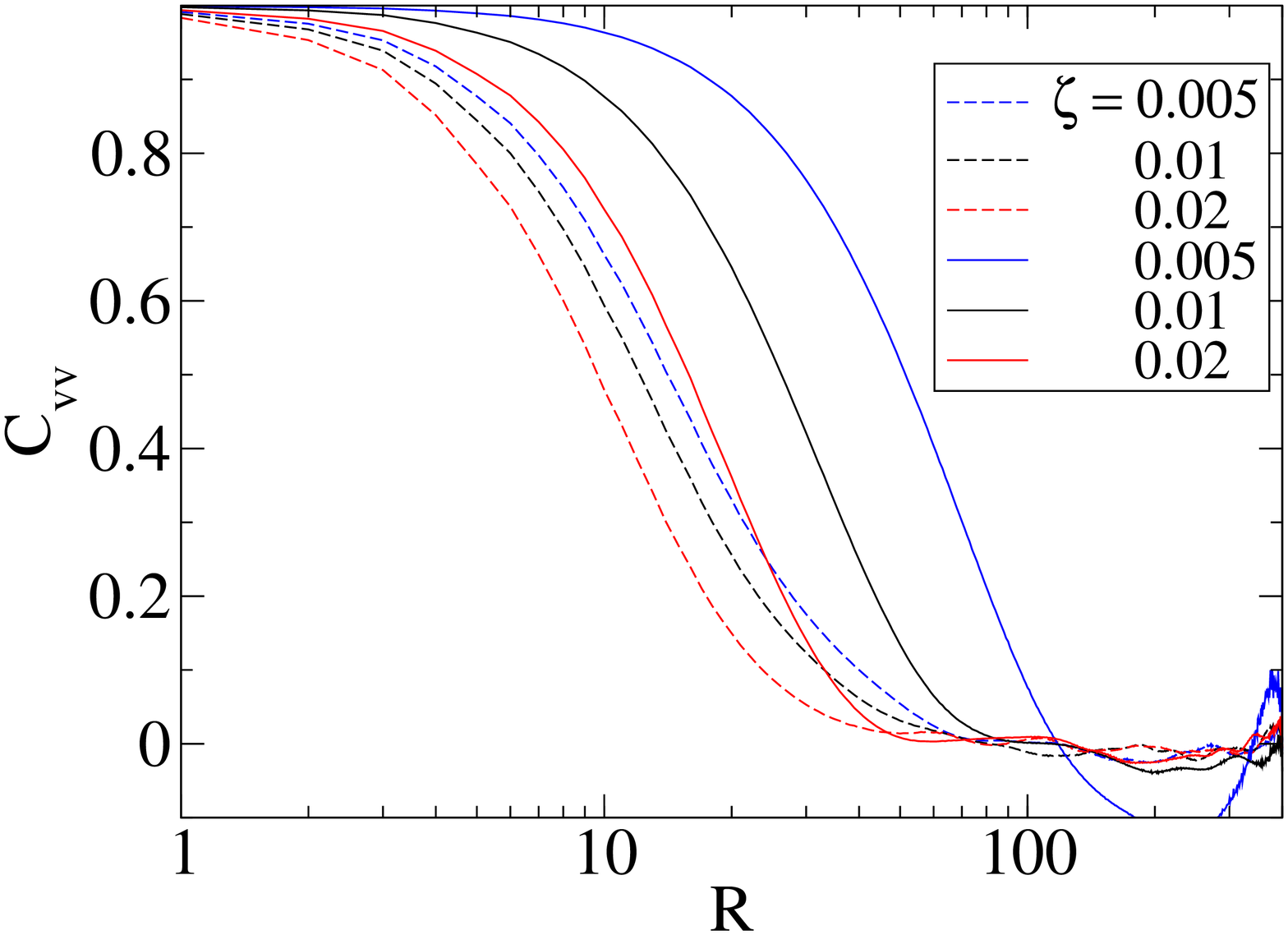}}
\subfigure[]{\includegraphics[trim = 0 0 0 0, clip, width=0.48\linewidth]{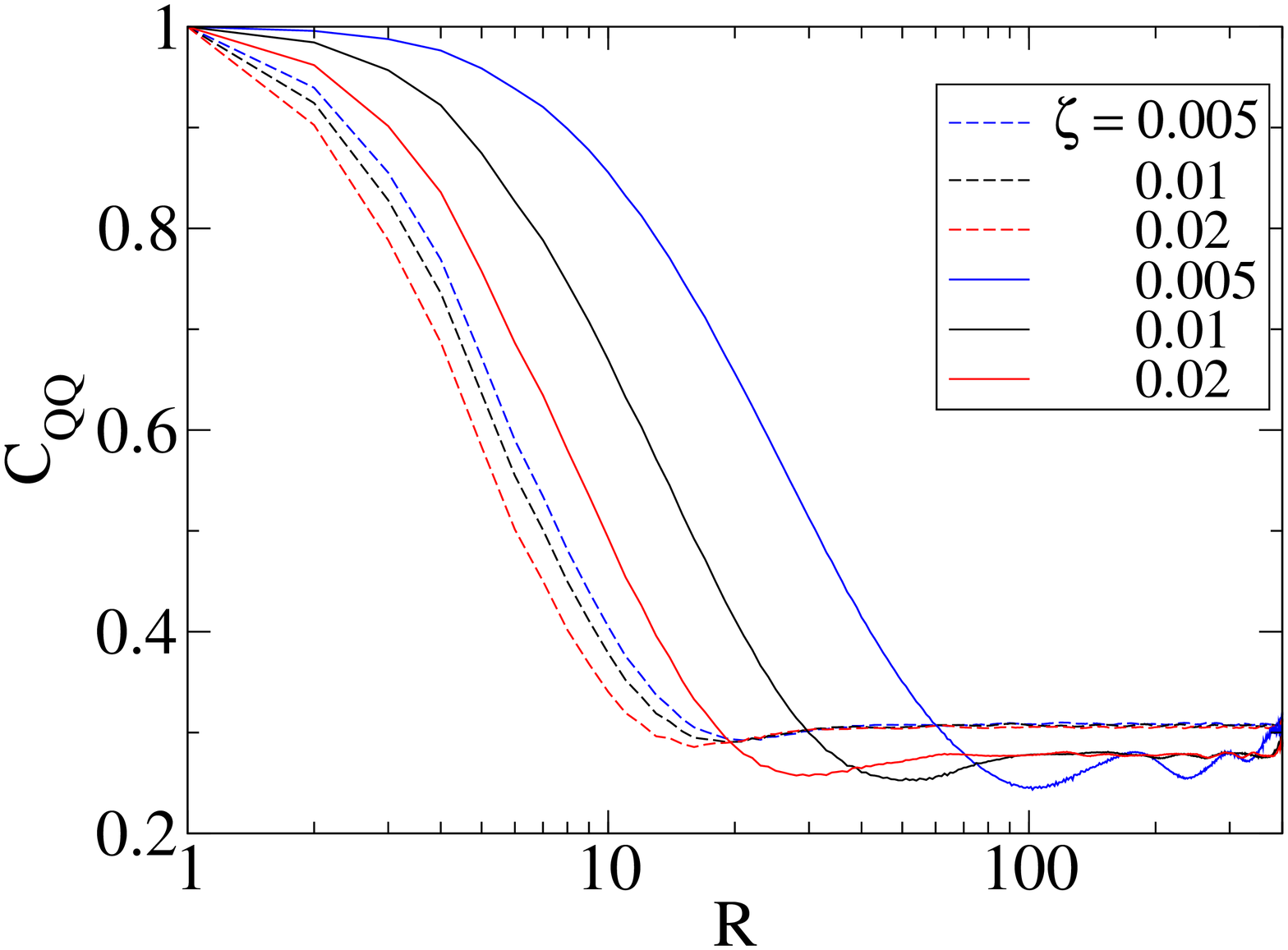}}
\caption{Activity, $\zeta$, dependence of (a) the velocity correlation functions, (b) the order parameter correlation function at $T < T_{\textnormal{IN}}$ (dashed lines), $T>T_{\textnormal{IN}}$ (continuous lines).}
\label{fig:corrs}
\end{figure}

To understand the role of intrinsic free energy in determining the characteristic length scales of the active turbulence state we measured the velocity correlation functions $C_{vv}(R) = \langle \mathbf{v} (R){\cdot}\mathbf{v}(0) \rangle / \langle\mathbf{v} (0)^2\rangle$ and order parameter correlation functions $C_{QQ}(R) = \left(\langle\mathbf{Q}(R){:}\mathbf{Q}(0)\rangle{-}\langle\mathbf{Q}(\infty){:}\mathbf{Q}(0)\rangle\right){/}\left(\langle \mathbf{Q}(0)^2\rangle{-}\langle\mathbf{Q} (\infty){:}\mathbf{Q}(0)\rangle\right)$, at two different temperatures $T < T_{\textnormal{IN}}$ (Fig.~\ref{fig:corrs}: dashed lines) and $T>T_{\textnormal{IN}}$ (Fig.~\ref{fig:corrs}: continuous lines). 
Deep in the nematic phase we find only weak dependence of velocity and order parameter correlation functions consistent with previous investigations \cite{Dogic2012, Aranson2012, Jorn2013, SumeshPRL2013, Giomi2014turb,ourpta2014}. However, near $T_{\textnormal{IN}}$, change in activity changes the degree of nematic ordering and in response, all characteristics of the turbulent field show a stronger dependence on activity. This qualitatively different behaviour might help to establish the role of activity driven ordering in experiments.

\section{Concentration segregation} In the context of dry active systems (no hydrodynamics) it is customary to add an active current to the Cahn-Hilliard equation,  $\zeta_{\phi}\nabla \cdot Q$, where $\zeta_{\phi}$ is a constant \cite{Narayan2007, Marchetti2013}. This accounts for the curvature in the nematic order inducing a local polarity, and produces phase separation \cite{Chate2013,Chate2014}. We obtain concentration ordering without the addition of this term or any free energy terms.

We have seen that active forces set up large scale flows. Advection of concentration by these flow fields is the only source term in Eq.~\ref{eqn:conc} and this generates concentration ordering as seen in Fig.~\ref{fig:turb}. For gradients to form the time scale of the active forcing, $\tau_{\zeta}=\eta/\zeta$, must be smaller than the relaxation times of the nematic order, $\tau_{Q}=L^{2}/\Gamma_Q K_Q$, and concentration, $\tau_{\phi}=L^{2}/\Gamma_{\phi} K_{\phi}$ \cite{Mahadevan2011}. 

\section{Discussion} To summarise, we demonstrate that, even in the absence of free energy contributions, activity-induced flow results in nematic ordering and concentration segregation in active nematics. We show that the active contribution to the hydrodynamics in systems of active rods can be written as an effective molecular field of Landau-de Gennes form. This is different to the active term in the free energy that was proposed in the generalised force-flux relations of active gels \cite{Kruse2004,Marchetti2013}. Our results also have consequences in various aspects of active systems and we conclude by discussing these:

\textit{Link with fluid dynamical approaches:} Numerical simulations and linear stability analysis of suspensions of extensile microswimmers (pushers)  have shown that the isotropic state is unstable to fluctuations which drive the system into an ordered state \cite{Shelley2008, Shelley2008b, Ganesh2009}. This instability was not observed in case of contractile microswimmers (pullers) with dipolar hydrodynamic interactions although incorporating steric effects has been shown to result in local order \cite{Ezhilan2013,Ishikawa2008,Alarcon2013,Zottl2014}. The physical mechanism elucidated here naturally explains these results, and provides a link between active liquid crystal theory and the fluid mechanical framework used to describe swimmer suspensions. It is interesting to note that segregation in the concentration of active extensile particles has been seen in numerical simulations \cite{Shelley2008b}. However, a linear stability analysis for isotropic suspensions did not predict any concentration segregation, which indicated the role of coupling between flow and concentration fields in segregating $\phi$.

\textit{Effective temperature:} Despite the complex non-equilibrium behaviour, it is possible to find an effective equilibrium description for some aspects of active systems, such as ordering and condensation transitions \cite{Tailleur2015}. A particularly appealing concept of effective temperature has been suggested as a possible way to represent the (kinetic) strength of the activity of particles \cite{Palacci2010}, and its generic applicability has been recently examined \cite{Szamel2014}. The representation of activity as an enhanced effective temperature can be rigourously derived for dilute suspensions of active particles provided they do not have any tendency towards orientational ordering \cite{Ramin2012}. However this does not apply to swimming bacteria due to the asymmetry in their shapes and the flow fields they generate. Moreover, even spherical Janus particles with catalytic self-propulsion have been recently shown to have asymmetric velocity profiles due to closed proton current loops near their catalytic cap \cite{Ebbens2014, Aidan2014} that could lead to a net extensile flow field \cite{Aidan2014arxiv}.

Recall the phase diagram (Fig.~\ref{fig:phase}) that demarcates the equilibrium vs non-equilibrium forces in active nematics and the asymmetry in the behaviour of extensile and contractile systems with $\zeta$. This highlights the fact that positional and orientational degrees of freedom respond differently to the active stress: the flow--induced enhanced diffusion is proportional to $\zeta^2$, whereas the orientational response is predominantly proportional to $\zeta$ and differs for extensile and contractile systems. Therefore, our results suggest that all consequences of activity cannot be represented by a single effective temperature when the active elements are asymmetric, which is most commonly the case in active matter. Therefore our analysis cautions against the emerging idea of interpreting activity as an effective temperature in describing the dynamics of active systems.

\textit{Reduced description:} Our analysis leads to a system of equations that are independent of widely used, but somewhat arbitrary, material parameters in the free energy. To obtain a minimal set of dimensionless variables we note that our numerical simulations show that $\boldmath{\Pi}^{passive}$ does not change the qualitative behaviour shown by active nematics and can be dropped. We choose two length scales, $L$, $L_D$ as the characteristic length scales for flow and diffusion, and $U$ as the characteristic velocity to non-dimensionalise the equations. Thus, along with the equation of continuity (Eq.~\ref{eqn:cont}), we have
\begin{align}
Re (\partial_t + u_k \partial_k) u_i &= \partial_j (-P \delta_{ij}+2 E_{ij} - Z \phi Q_{ij}),
\label{eqn:rns} \\
(\partial_t + u_k \partial_k) Q_{ij} - S_{ij} &= Pe_Q^{-1} \nabla^{2}Q_{ij},
\label{eqn:lrc}\\
\partial_{t}\phi+\partial_{i}(u_{i}\phi)&=Pe_{\phi}^{-1} \nabla^{2}\phi.
\label{eqn:rconc}
\end{align}
where the Reynolds number $Re = \frac{LU\rho}{\eta}$ describes the ratio of inertial to viscous forces, $Z = \frac{\zeta L}{ \eta U}$ describes the ratio of active to viscous forces, $Pe_Q = \frac{L_D^2 U}{\Gamma_Q L}$ and $Pe_{\phi} = \frac{L_D^2 U}{\Gamma_{\phi} L}$ respectively describe the advective to diffusive transport rate of $Q$ and $\phi$. In addition, we will have the alignment parameter $\lambda$. These equations are independent of all other material parameters and represent a simpler system than equations used in the literature \cite{Mahadevan2011, Giomi2013, SumeshPRL2013, MatthewPRL,Giomi2014turb} to describe the dynamics of extensile active nematics.

\begin{acknowledgments}
We acknowledge funding from the ERC Advanced Grant MiCE.
\end{acknowledgments}

\bibliography{refe.bib}

\end{document}